# Using Perturbations to Probe the Neural Control of Rhythmic Movements

Tim Kiemel[1], David Logan[1], John J. Jeka[2,3]
[1]*Department of Kinesiology, University of Maryland,* [2]*Departments of Kinesiology and* [3]*Bioengineering Temple University*

**Correspondence:**
Tim Kiemel
Department of Kinesiology
University of Maryland
College Park, MD  20742
phone: 301-405-2488
email: kiemel@umd.edu

**Abstract**

Small continuous sensory and mechanical perturbations are often used to identify properties of the closed-loop neural control of posture and other systems that are approximately linear time invariant. Here we extend this approach to study the neural control of rhythmic behaviors such as walking. Our method is based on the theory of linear time periodic systems, with modifications to account for ability of perturbations to reset the phase of a rhythmic behavior. We characterize responses to perturbations in the frequency domain using harmonic transfer functions and then convert to the time domain to obtain phase-dependent impulse response functions ($\phi$IRFs) that describe the response to a small brief perturbation at any phase of the rhythmic behavior. The $\phi$IRFs describing responses of kinematic variables and muscle activations (measured by electromyography) to sensory and mechanical perturbations can be used to infer properties of the plant, the mapping from muscle activation to movement, and of neural feedback, the mapping from movement to muscle activation. We illustrate our method by applying it to simulated data from a model and experimental data of subjects walking on a treadmill perturbed by movement of the visual scene.

# 1. Introduction

To produce stable walking, the nervous system must perform a task of astounding complexity. Using sensory information about movement of the body's multiple mechanical degrees of freedom, the nervous system coordinates the activation of dozens of muscles. From a control theory perspective, these behaviors involve the closed-loop interaction between the plant $P$, the entity being controlled, and neural feedback $F$ (Fig. 1A). The plant describes how the control signal $u(t)$ that specifies muscle activation, as measured by electromyography (EMG), causes movement $y(t)$; neural feedback describes how movement causes muscle activation via sensory inputs that are integrated by our nervous system. The plant, along with task constraints, defines the control problem that confronts the nervous system. Neural feedback is the nervous system's solution to this control problem. Thus, identifying properties of the plant and neural feedback yields insight into the neural control of walking.

The plant $P$ and neural feedback $F$ are *open-loop* mappings. Fig. 1A shows two additional open-loop mappings: the direct sensory effect $S$ and the direct mechanical effect $M$. Neural feedback $F$ describes how movement $y(t)$ would affect the control signal $u(t)$ in a hypothetical open-loop condition in which the control signal does not affect movement (Fig. 1B). The direct sensory effect $S$ describes how a sensory perturbation $v(t)$ would affect the control signal in this open-loop condition. Similarly, the plant $P$ describes how the control signal $u(t)$ would affect movement $y(t)$ in a hypothetical open-loop condition in which movement does not affect the control signal (not shown). The direct mechanical effect $M$ describe how a mechanical perturbation $d(t)$ would affect movement in this open-loop condition.

For a neuromechanical system, such as the neural control of standing, that is approximately linear time invariant (LTI), frequency response functions (FRFs) are used to characterize open-loop mappings in the frequency domain [1]. Let $U(f)$, $Y(f)$, $V(f)$, and $D(f)$ be the Fourier transforms of $u(t)$, $y(t)$, $v(t)$ and $d(t)$, respectively, where $f$ is frequency. Also, let $H_P(f)$, $H_F(f)$, $H_S(f)$ and $H_M(f)$ be the FRFs of $P$, $F$, $S$ and $M$, respectively. Then the neuromechanical system of Fig. 1 can be written as

$$Y(f) = H_P(f)U(f) + H_M(f)D(f), \tag{1a}$$
$$U(f) = H_F(f)Y(f) + H_S(f)V(f). \tag{1b}$$

For an LTI system, FRFs are also used to characterize closed-loop responses. We will use $H_{ab}(f)$ to denote the closed-loop FRF from $a(t)$ to $b(t)$. If all perturbations are mutually independent, from (1) we have the following relationships among kinematic and EMG closed-loop responses to a sensory perturbation $v(t)$ and a mechanical perturbation $d(t)$:

$$H_{vy}(f) = H_P(f)H_{vu}(f), \tag{2a}$$
$$H_{du}(f) = H_F(f)H_{dy}(f), \tag{2b}$$
$$H_{vu}(f) = H_F(f)H_{vy}(f) + H_S(f), \tag{2c}$$
$$H_{dy}(f) = H_P(f)H_{du}(f) + H_M(f). \tag{2d}$$

From (2a) we have that for a sensory perturbation $v(t)$, the plant $P$ maps EMG responses to kinematic responses. Similarly, from (2b) we have that for a mechanical perturbation $d(t)$, neural feedback $F$ maps kinematic responses to EMG responses. These two statements are the basis of the joint input-output (JIO) method of closed-loop system identification for LTI systems [2,3], which has been applied to the postural control of upright stance (e.g., [4–7]). The JIO method uses closed-loop kinematic and EMG responses to infer properties of the plant and neural feedback. The frequency-domain relationships of (2) have analogous time-domain descriptions:

$$h_{vy} = h_P * h_{vu}, \tag{3a}$$
$$h_{du} = h_F * h_{dy}, \tag{3b}$$
$$h_{vu} = h_F * h_{vy} + h_S, \tag{3c}$$
$$h_{dy} = h_P * h_{du} + h_M, \tag{3d}$$

where lowercase $h$ denotes an impulse response function (IRF) and '*' denotes convolution.

In this paper we extend the relationships of (3) to the neural control of rhythmic movements such as walking. We idealize rhythmic movement as the output of a neuromechanical closed-loop system (Fig. 1) with a stable limit cycle. The response of such a limit-cycle system to a transient perturbation will, in general, have two components: a transient response and a phase shift that persists after the transient response decays away. By comparison, the response of a stable linear time periodic (LTP) system to a transient perturbation has only the transient component. We will use LTP theory as part of our extension of (2) from LTI systems to limit-cycle systems. Standing can be approximated as a stable LTI system (e.g., [3,8]), walking in sync with a metronome can be approximated as a stable LTP system, and steady-state walking in general can be approximated as a stable limit-cycle system (e.g., [9,10]). Our extension here of LTI and LTP methods to limit-cycle systems provides an unified approach to non-parametrically describing these three types of systems and allows investigation of the neural control of walking using extensions of powerful methods that have been successfully applied to standing.

An extension of the FRF exists for LTP systems called the *harmonic transfer function* (HTF) [11–15], which can be used to characterize both open- and closed-loop mappings in the frequency domain. Recently, Ankarali and Cowan [16] developed a new system identification method that uses HTFs to describe the responses of hybrid locomotor systems to perturbations.

The time-domain characterization of a LTP mapping is the phase-dependent impulse response function [14,15], which we will refer to as a $\phi$IRF ($\phi$ denotes phase) to distinguish it from the IRF for an LTI system. The effect of a perturbation $v(t)$ on the variable $u(t)$ is given by the $\phi$IRF $h_{vu}(t_r, t_p)$:

$$u(t_r) - u_0(t_r) = \int_{-\infty}^{t_r} h_{vu}(t_r, t_p) v(t_p) dt_p, \tag{4}$$

where $u_0(t_r)$ is the unperturbed $T$-periodic output of the LTP system. The $\phi$IRF $h_{vu}(t_r, t_s)$ describes the response at time $t_r$ to an impulse perturbation applied at time $t_p$. We have that $h_{vu}(t_r, t_p) = 0$ for $t_r < t_p$ and $h_{vu}(t_r + T, t_p + T) = h_{vu}(t_r, t_p)$.

Suppose that the output $u(t) - u_0(t)$ of (4) is the input to a second LTP mapping whose output is $x(t) - x_0(t)$ and whose $\phi$IRF is $h_{ux}$. Then the mapping from $v(t)$ to $x(t) - x_0(t)$ has $\phi$IRF $h_{vx} = h_{ux} * h_{vu}$, where

$$(h_{ux} * h_{vu})(t_r, t_p) = \int_{t_p}^{t_r} h_{ux}(t_r, t) h_{vu}(t, t_p) dt \tag{5}$$

defines convolution for $\phi$IRFs. Now the time-domain description (3) of an LTI system also serves as a description of an LTP system. The key observation is that (3) is a good approximation of the response of a limit-cycle system to a small transient perturbation.

It may seem more natural to extend the frequency-domain description (2) from LTI to limit-cycle systems, since this would involve multiplication of HTFs rather than convolutions of $\phi$IRFs. However, as we will see, it is problematic to use HTFs for limit-cycle systems because closed-loop HTFs are unbounded due to phase resetting. For that reason, we use the time-domain description (3).

Although we characterize limit-cycle mappings in the time domain, it is not efficient to directly estimate $\phi$IRFs from data (see *Discussion*). Instead, we estimate phase and use estimated phase rather than time as the independent variable. Doing so allows us to use LTP techniques to characterize the input-output mapping in the frequency domain using harmonic transfer functions (HTFs). We then convert the HTFs to $\phi$IRFs in the time domain. The key to our approach is that we also perform an LTP analysis of estimated phase in order to produce an estimate of the $\phi$IRF that, to first order in the perturbation size, does not depend on the particular method used to estimate phase.

The paper is organized as follows. In section 2 we describe our method of computing a $\phi$IRF from experimental data in which a small broadband perturbation is applied to rhythmic movement. Technical details are in the Appendix. In section 3, we apply our method to simulated data from a simple oscillator model. In section 4, we apply our method to experimental walking data. Finally, in section 5 we discuss how our method can be used to infer mechanisms underlying the control of rhythmic movement.

## 2. Method of computing $\phi$IRF

We consider the effect of a small continuous broadband perturbation, $u(t)$, on a response variable, $y(t)$, describing some aspect of a rhythmic movement. In our experimental example of section 4, $u(t)$ is the velocity of a virtual visual scene surrounding a person walking on a treadmill (Fig. 2A) and $y(t)$ is a rectified electromyographic (EMG) signal recorded from a muscle (Fig. 2B) or a kinematic variable. For each experimental subject, we have on the order of $10^3$ to $10^4$ total cycles of activity from multiple trials of walking. Our goal is to compute the $\phi$IRF that describes the effect of $u(t)$ on $y(t)$. In this section we describe the steps in our method of computing the $\phi$IRF. See the appendix for justification of our method and additional details.

To perform our analysis we need a third signal, $\theta(t)$, that estimates the phase of the rhythmic movement. We want $\theta(t)$ to be causal, continuously differentiable, and monotonically increasing. We also require that if the rhythmic movement is strictly periodic, then $\theta(t) = f_0 t$, where $f_0$ is the movement's frequency, and integer values of $\theta(t)$ occur at the times $t_k$ ($k = 1, \ldots, K$) of some event that happens once for cycle. In other words, to call $\theta(t)$ an estimate of phase we require that it equals actual phase if there is no cycle-to-cycle variability.

For walking experimental data, the event times $t_k$ are the times at which the heel of a given leg first strikes the support surface during each gait cycle. To compute $\theta(t)$ we first compute a discontinuous causal estimate of phase as $\theta_d(t) = k + \hat{f}_0(t - t_k)$ for $t_k \leq t < t_{k+1}$, where $\hat{f}_0 = 1/\bar{T}$ and $\bar{T}$ is the mean of the cycle periods $t_{k+1} - t_k$ ($k = 1, \ldots, K-1$). The computation of $\theta_d(t)$ is shown in Fig. 2C for a hypothetical example with exaggerated variability in cycle periods and in Fig. 2D for actual experimental data (blue line segments). Our continuously-differentiable causal estimate of phase, $\theta(t)$, is obtained by applying a second-order low-pass filter to $\theta_d(t)$:

$$\ddot{\theta}(t) + 2\rho(\dot{\theta}(t) - \hat{f}_0) + \rho^2 \theta(t) = \rho^2 \theta_d(t), \tag{6}$$

where the parameter $\rho$ is chosen so that $\theta(t)$ is monotonically increasing (red curves in Fig. 2C–D). From the phase estimate $\theta(t)$, we compute its time derivative $\dot{\theta}(t)$ (Fig. 2E), which will be used in our method of computing the $\phi$IRF.

Now that we have our perturbation signal $u(t)$, our response variable $y(t)$, and the time derivative $\dot{\theta}(t)$ of our phase estimate, we switch the independent variable from time to estimated phase so that we can perform an LTP analysis of our signals. We use the symbol $\vartheta$ to denote approximate phase as an independent variable to distinguish it from $\theta$, which is the function that maps time $t$ to approximate phase. Since $\theta(t)$ is a monotonically increasing function of time $t$, we can define its inverse $p$: $p(\theta(t)) = t$ and $\theta(p(\vartheta)) = \vartheta$. Now we can define the following signals with $\vartheta$ as the independent variable: the perturbation signal, $\tilde{u}(\vartheta) = u(p(\vartheta))$; the response variable, $\tilde{y}(\vartheta) = y(p(\vartheta))$; and the time derivative of the phase estimate, $\tilde{q}(\vartheta) = \dot{\theta}(p(\vartheta))$. Fig. 2F shows $\tilde{y}(\vartheta)$ and Fig. 2G shows the mean periodic waveform $y_0(\vartheta)$ obtained by averaging $\tilde{y}(\vartheta)$ across the cycles in the trial.

The last row of Fig. 2 shows the signals that are used in our LTP analysis: $\tilde{u}(\vartheta)$, the perturbation signal; $\tilde{y}^{(1)}(\vartheta) = \tilde{y}(\vartheta) - y_0(\vartheta)$, the response variable minus its mean periodic waveform; and $\tilde{q}^{(1)}(\vartheta) = \tilde{q}(\vartheta) - \hat{f}_0$, the time derivative of estimated phase minus estimated cycle frequency. From our LTP analysis (see Appendix) we obtain the $\phi$IRF $\tilde{h}_y(\vartheta_r, \vartheta_p)$ from $u(\vartheta)$ to $\tilde{y}^{(1)}(\vartheta)$ and the $\phi$IRF $\tilde{h}_q(\vartheta_r, \vartheta_p)$ from $u(\vartheta)$ to $\tilde{q}^{(1)}(\vartheta)$.

The final steps in our method compensate for the switch from time to estimated phase as the independent variable. See the Appendix for justification. Using (15) in the Appendix, we integrate $\tilde{h}_q(\vartheta_r, \vartheta_p)$ with respect to $\vartheta_r$ to obtain the $\phi$IRF $h_\theta(t_r, t_p)$ from $u_1^{(1)}(t)$ to $\theta^{(1)}(t)$. Then we obtain the desired $\phi$IRF from $u(t)$ to $y(t)$ as

$$h_y(t_r, t_p) = \tilde{h}_y(t_r, t_p) + y_0'(t_r) h_\theta(t_r, t_p). \tag{7}$$

## 3. Illustration of the method with simulated data

### 3.1. Estimation of closed-loop $\phi$IRFs

To illustrate our method of estimating the $\phi$IRF, we consider a control model with a LTI plant that maps the control signal $u(t)$ to the kinematic signal $y(t)$:

$$\ddot{y}(t) + a\dot{y}(t) + \omega_0^2 y(t) = a\omega_0[u(t) + d(t)], \tag{8a}$$

where $d(t)$ is a known external mechanical perturbation. Feedback maps $y(t)$ to $u(t)$:

$$\dot{\phi}(t) = \omega_0 + a_{\mathrm{f}}[\omega_0 z(t)\cos\phi(t) - \dot{z}(t)\sin\phi(t)], \quad z(t) = y(t) - v(t) + w(t), \tag{8b}$$

$$\ddot{\phi}_{\mathrm{f}}(t) + 2r_{\mathrm{f}}\bigl[\dot{\phi}_{\mathrm{f}}(t) - \omega_0\bigr] + r_{\mathrm{f}}^2\phi_{\mathrm{f}}(t) = r_{\mathrm{f}}^2\phi(t), \quad u(t) = \cos\phi_{\mathrm{f}}(t), \tag{8c}$$

where $v(t)$ is a known external sensory perturbation and $w(t)$ is sensory noise. In the absence of perturbations and noise, $y(t) = y_0(t) \triangleq \sin\omega_0 t$ and $u(t) = u_0(t) \triangleq \cos\omega_0 t$ is a solution of the model.

We consider two methods of computing estimated phase $\theta(t)$. In the first method, estimated phase $\theta(t)$ is computed directly from phase $\phi(t)$ using a second-order low-pass filter:

$$\ddot{\theta}(t) + 2\rho\bigl(\dot{\theta}(t) - f_0\bigr) + \rho^2\theta(t) = \rho^2\phi(t),$$

where $f_0 = \omega_0/2\pi$. Note that this method is not feasible for experimental data, since one cannot directly measure phase. The second method is the one used for experimental data described in section 2, except that event times are those times at which $y(t)$ passes upward through 0.

Figure 3 shows the results of applying our method of computing $\phi$IRFs to simulated data from the model. Each subplot shows an $\phi$IRF. The horizontal axis indicates normalized perturbation time $t_{\mathrm{p}}$, which is equivalent to perturbation phase. The vertical axis indicates normalized response time $t_{\mathrm{r}}$. For each perturbation phase $t_{\mathrm{p}}$ and normalized response time $t_{\mathrm{r}}$, color indicates the $\phi$IRF value: green for 0, red for positive, and blue for negative. $\phi$IRFs in the top row were computed using estimated phase obtained by directly filtering phase in the model. The $\phi$IRFs in the bottom row were computed using estimated phase based on phase-0 times.

In the first column of Fig. 3 are estimated $\phi$IRFs $h_\theta(t_{\mathrm{r}}, t_{\mathrm{p}})$ from the perturbation $v(t)$ to estimated phase $\theta(t)$. Note that the $\phi$IRF based on event times (Fig. 3E) rises more slowly than the $\phi$IRF based on filtered phase (Fig. 3A). This is due both to the low-pass filtering used to produce $\phi_{\mathrm{f}}(t)$ and $\rho_{\mathrm{f}}(t)$ in the model and the fact that estimated phase does not reflect the effect of a perturbation until the next phase-0 event occurs.

In the second column of Fig. 3 are $\phi$IRFs $y_0'(t_{\mathrm{r}})h_\theta(t_{\mathrm{r}}, t_{\mathrm{p}})$ describing the component of the $\phi$IRF of $y$ due the effect of the perturbation on estimated phase. The estimate of the unperturbed sinusoidal waveform $y_0(t)$ is similar for both methods of estimating phase (not shown), so differences in the phase $\phi$IRF of $y$ between Fig. 3B and Fig. 3F are almost entirely due to differences in the $\phi$IRFs of phase in Fig. 3A and Fig. 3E. In the third column of Fig. 3 are $\tilde{h}_y(t_{\mathrm{r}}, t_{\mathrm{p}})$, the transient $\phi$IRF of $y$. The transient $\phi$IRF of $y$ is the component of the $\phi$IRF of $y$ that is not due to the effect of the perturbation on estimated phase.

In the last column of Fig. 3 are estimates of the $\phi$IRF of $y$, computed as the sum of the phase $\phi$IRF of $y$ in column 2 and the transient $\phi$IRF of $y$ in column 3. Note that both methods of estimating phase produce $\phi$IRFs of $y$ that are indistinguishable from each other, demonstrating that our method works as intended for the given simulated data. The similarity of the both $\phi$IRFs is further illustrated in Fig. 4, which shows vertical slices through the $\phi$IRFs in Fig. 3 at a perturbation phase of 0. The upper limit of normalized response time $t_r$ has been increased to illustrate that both $\phi$IRFs of phase in Fig. 4A converge to the same value for large $t_r$. This asymptotic value is the eventual phase shift produced by a perturbation at the given perturbation phase.

### 3.2. Relationships among open- and closed-loop $\phi$IRFs.

Our primary motivation to measure closed-loop responses of a neuromechanical control system is to provide insight into the mechanisms underlying the system's behavior. The system is viewed as a set of interacting components with each component described by the open-loop mapping from its inputs to its outputs [17] (Fig. 1A). For example, the plant is an open-loop mapping from control signals to kinematics and feedback is the open-loop mapping from kinematics to control signals. The goal is to understand the system's closed-loop behavior based on understanding the properties of each open-loop component.

We use $\phi$IRFs to characterize open-loop mappings, allowing us to use these open-loop $\phi$IRFs to predict closed-loop $\phi$IRFs. As an example, consider opening the loop in model (8) by removing the effect of the control signal $u(t)$ on kinematics $y(t)$ (Fig. 1B). We do this by letting $y(t) = y_0(t) + y_p(t)$ in (8b), where $y_0(t) = \sin \omega_0 t$ is the periodic kinematic waveform of the unperturbed system and $y_p(t)$ is a small specified kinematic perturbation. Now we can define open-loop mappings for neural feedback $F$ and the direct sensory effect $S$ from the kinematic perturbation $y_p(t)$ and sensory perturbation $v(t)$, respectively, to the control-signal deviation $u(t) - u_0(t)$, where $u_0(t) = \cos \omega_0 t$ is the unperturbed control-signal waveform. These mappings are approximately LTP with time as the independent variable. There is no need to estimate phase, because the periodic component $y_0(t)$ of $y(t)$ acts as a perfect clock that prevents phase resetting. Therefore, we can compute open-loop $\phi$IRFs $h_F$ and $h_S$ simply by computing the HTF in the frequency domain and converting to the time domain (Fig. 5B and C). Note that $h_S = -h_F$ in our model.

Similarly, we can open the loop by letting $u(t) = u_0(t) + u_p(t)$ in (8a), where $u_p(t)$ is a small specified control-signal perturbation. Then the plant $P$ and direct mechanical effect $M$ and are defined as the open-loop mappings from the control-signal perturbation $u_p(t)$ and mechanical perturbation $d(t)$, respectively, to the kinematic deviation $y(t) - y_0(t)$. The corresponding $\phi$IRFs $h_P$ and $h_M$ are shown in Fig. 5A and D. Note $h_M = h_P$ in our model.

Having defined the open-loop mappings $P$, $F$, $S$ and $M$ (Fig. 1A) and computed their $\phi$IRFs for our model (Fig. 5A–D), we can now use (3) to compute closed-loop $\phi$IRFs that describe how the closed-loop system responds to external perturbations (Fig. 5E–H). The $\phi$IRFs $h_{vy}$ and $h_{vu}$ describing kinematic and control-signal responses to the sensory perturbation were computed by solving the coupled equations (3a) and (3c), starting with $h_{vy}(t_p, t_p) = h_{vu}(t_p, t_p) = 0$ and using that $h_{vy}(t_r, t_p)$ only depends on $h_{vu}(t, t_p)$ for $t < t_r$ and *vice versa*. Similarly, the $\phi$IRFs $h_{dy}$ and $h_{du}$ describing kinematic and control-signal responses to the mechanical perturbation were computed by solving the coupled equations (3b) and (3d).

### 4. Illustration of the method with experimental walking data

We applied our method of computing $\phi$IRFs to subjects walking on a treadmill perturbed by movement of a virtual visual scene. The Institutional Review Board at the University of Maryland approved the procedures used in the experiment. We briefly describe the experimental methods here; see [8] for additional details, where similar experimental methods were used to study upright stance. Data were collected from 20 subjects walking on a treadmill at 4.99 km/hr (1.39 m/s). Each subject walked surrounded by three screens (one in front and one on each side) displaying a random pattern of triangles representing a virtual visual scene consisting of three walls. The visual scene rotated about a fixed medial-lateral axis whose anterior-posterior position roughly matched that of the subject and whose vertical position above the surface matched the ankle height of the subject. Rotation signals were produced by passing white noise through a first-order low-pass Butterworth filter with a cutoff of 0.02 Hz and an eighth-order low-pass Butterworth filter with a cutoff frequency of 5 Hz. We used the angular velocity of the visual-scene rotation as the perturbation signal, which had a root-mean-square value of 6.9 deg/s. Due to the filtering described above, visual-scene angular velocity was 0.02–5 Hz bandpass-filtered white noise. A positive value of the perturbation signal corresponds to a forward rotation of the visual scene. We recorded surface electromyographic (EMG) signals from various muscles and the locations of various kinematic markers placed on the subject's body. EMG signals indicate the timing and relative level of muscle activation by the nervous system. To obtain estimated phase, we computed the times at which the heel of a given leg landed on the treadmill surface. These heel strike times indicate the end of the leg's swing phase and the beginning of the leg's stance phase. We estimated phase from the heel-strike times using the same method we applied to the simulated data in section 4.

Figure 6 shows mean $\phi$IRFs describing responses to visual-scene velocity. Above each $\phi$IRF is the mean waveform of the response signal as a function of the phase of the gait cycle. For kinematic responses, we will restrict our attention to changes in walking speed (Fig. 6O) and the resulting changes in the subject's position on the treadmill (Fig. 6P). When the visual-scene moved forward, walking speed increased after some delay, as indicated by the change from green to red with increasing normalized response time in the $\phi$IRF of Fig. 6O. (Equivalently, when the visual-scene moved backward, walking speed decreased.) This increase in speed led to a prolonged forward change in the subject's position (Fig. 6P). Our primary goal is to understand how changes in muscle activation by the nervous system led to the change in speed.

The first muscle we will consider is the tibialis anterior (TA) muscle, which acts to dorsiflex the ankle, that is, bring the shin and toes closer together. The mean TA EMG waveform has a peak in activity near the beginning of the gait cycle (phase 0) when heel strike occurs (Fig. 6I). The $\phi$IRF in Fig. 6I shows changes from this mean waveform in response to movement of the visual scene. The red area indicated by the arrow represents a response in which forward visual-scene velocity led to an increase in TA activity. The phase dependency of this response is indicated by the horizontal axis, with the white bar indicating the swing phase of gait (phases -0.38 to 0) and the black bar indicating the stance phase of gait (phases 0 to 0.62). In this case, the red area lies above the middle of the white bar, indicating that visual-scene motion during mid-swing led to a change in TA activity. The timing of the response is indicated by the vertical axis. In this case, the red area lies to the right of the lower end of the first black bar, indicating that the response occurred during early stance, when the mean TA activity is high.  In summary, the red area indicates that when the visual scene moved forward during mid-swing, TA activity increased early in the following stance phase. Similar early-stance responses were seen in quadriceps muscles that extend the knee, the vastus lateralis, vastus medialis and rectus femoris muscles (black arrows in Fig. 6J–L), suggesting that these muscles work in concert with the TA muscle to increase walking speed when the visual scene moves forward during midswing.

In addition to early-stance responses, there were mid- to late-stance responses in the soleus, medial and lateral gastrocnemius muscles (Fig. 6A–C).  These muscles act to plantarflex the ankle, that is, move the shin and toes further apart. The plantarflexors are thought to contribute to forward propulsion and, thus, changes in speed.   When the visual scene moved forward during swing or early stance, the activity of one or more of the plantarflexors increased during late stance. (There was also some decreased activity in mid-stance for reasons that are unclear.)  Taking into account that plantarflexors of the other leg are active half a cycle out of phase, our results indicate that forward visual-scene motion at any phase of the gait cycle leads to increased plantarflexor activity.

Along with EMG responses during stance, there were EMG responses during early swing.  For example, in addition to extending the knee, the rectus femoris muscle also acts to flex the hip.  This biarticular action is related to a second peak in the mean waveform of rectus femoris activity in early swing (Fig. 6L) that is not present in the waveforms of the vasti muscles (Fig. 6J–K).  The early-swing rectus femoris activity increased when the visual scene moved forward during the preceding stance phase, as indicated by the red arrow in the $\phi$IRF of Fig. 6L. Similar early-swing responses were seen in the tensor fasciae latae and sartorius muscles (red arrows in Fig. 6M–N), whose actions also include hip flexion.

So far we have classified EMG responses into three categories: early-stance responses of ankle dorsiflexor and knee extensor muscles, mid- to late-stance response of ankle plantarflexor muscles, and early swing responses of hip flexor muscles. The remaining muscles of Fig. 6D–H are posterior muscles whose actions including knee flexion, hip extension, and extension of the vertebral column.  Their responses may work in concert with one of our three categories and/or be related to trunk orientation responses.

## 5. Discussion

We have illustrated our method by applying it to simulated data from a model and experimental data of subjects walking on a treadmill perturbed by movement of the visual scene. We now describe how this approach informs us about the mechanisms underlying those and other rhythmic systems through joint input-output (JIO) inference and short-latency inference. An additional benefit of using  the approach is that it allows for efficient study of Limit Cycle (LC) systems with use of continuous perturbations.

### 5.1. Joint input-output inference

For a closed-loop neuromechanical system consisting of a plant and neural feedback that produces rhythmic movement, there are two approaches that can be used to infer open-loop properties, such as those of the plant and feedback, based on closed-loop responses. We will refer to these approaches as joint input-output (JIO) inference and short-latency inference. JIO inference is based on the JIO method of closed-loop system identification for LTI systems [2,3], which has been applied to the postural control of upright stance [4–7]. The idea behind the JIO method is that for a sensory perturbation, the kinematic and control-signal responses depend on both the plant and neural feedback, but the relationship between them only depends on the plant. In fact, the plant maps control-signal responses to kinematic responses. In other words, if we know the control-signal responses to a sensory perturbation and if we know the plant, then we can predict the kinematic responses. For a LTI system with a single control signal, the JIO method (barring certain degeneracies) identifies a complete non-parametric characterization of the plant in the frequency domain, namely, the frequency response function (FRF) of the plant. The plant FRF can then be converted to an IRF in the time domain. For a system with multiple control signals, the number of sensory perturbations must equal the number of control signals. Otherwise, the JIO method only partially identifies the plant. Analogous to the use of sensory perturbations to identify the plant, the JIO method uses kinematic and control-signal responses to mechanical perturbations to identify neural feedback. In this case, neural feedback maps kinematics responses to control-signal responses.

In this paper, we have shown that JIO inference applies to LC systems. Namely, if we know the open-loop $\phi$IRF of the plant (Fig. 5A) and the closed-loop $\phi$IRF describing how the control signal responds to a sensory perturbation (Fig. 5G), then we can predict the closed-loop $\phi$IRF describing how the kinematic signal responds to the sensory perturbation (Fig. 5E). In this form, JIO inference can be used to test models of the plant. Similarly, mechanical perturbations can be used to test models of feedback. We have not addressed how to non-parametrically identify the plant or neural feedback. To identify the plant we expect that, as for LTI systems, the number of sensory perturbations must equal the number of control signals and that some yet-to-be-determined degeneracies must be avoided. In addition, identification of the plant in the frequency domain would be complicated by our use of both transient and phase HTFs to characterize closed-loop responses.

### 5.2. Short-latency inference

As suggested by its name, short-latency inference uses short-latency closed-loop responses to infer short-latency open-loop properties. For example, the initial closed-loop response of the control signal to a sensory perturbation (Fig. 5G) is equal to the initial open-loop direct effect $S$ of the perturbation (Fig. 5C). Since $S$ depends on sensory and neural processes, $S$ will have some relationship to neural feedback $F$ (Fig. 5B). In our model, $S = -F$. Thus, the initial closed-loop control-signal response in Fig. 5G is the negative of the initial response of neural feedback $F$ in 5B. Similarly, forward movement of the visual scene during walking is thought to create a (sub-conscious) illusion in the subject that he or she is walking more slowly than desired. As a result, the direct effect $S$ should be qualitatively similar to how neural feedback $F$ responds to an actual undesired slowing in walking speed. Since the initial open-loop response of S equals the initial closed-loop EMG response, we can interpret walking EMG responses as follows. When there is an undesired slowing in walking speed, neural feedback acts to modulate muscle activations. The initial responses due to this neural feedback is the given by the initial responses of the $\phi$IRFs in

Fig. 6A–N, which describe how the amplitude and timing of the neural responses depend on the phase of the gait cycle at which the slowing in walking speed occurs.

To summarize, for a sensory perturbation short-latency inference uses EMG responses to infer short-latency properties both of the direct effect $S$ of the perturbation and of neural feedback $F$, where inferences about $F$ require some assumption about the relationship between $S$ and $F$. In contrast, JIO inference uses the relationship between EMG and kinematic responses to infer properties of the plant $P$. For a mechanical perturbation, the situation is analogous. Short-latency inference uses kinematic responses to infer short-latency properties both of the direct effect $M$ of the perturbation and of the plant $P$, where inferences about $P$ require some assumption about the relationship between $M$ and $P$. JIO inference uses the relationship between kinematic and EMG responses to infer properties of neural feedback $F$.

For both JIO and short-latency inference, it is critical that $\phi$IRFs are defined with time rather than estimated phase as the independent variable. For example, the response of phase to a sensory perturbation depends on $S$, $P$ and $F$. If $\phi$IRFs are defined as functions of phase, JIO inference fails because the relationship between EMG and kinematic $\phi$IRFs will depend on $S$, $P$ and $F$, rather than just $P$. Short-latency inference also fails if phase has a short-latency response to the perturbation. Therefore, since our method of estimating closed-loop $\phi$IRFs depends on estimating phase, it was essential that, to first order in the size of perturbations, the resulting $\phi$IRFs do not depend on the method used to estimate phase. An alternative approach to studying neuromechanical control systems is to achieve a good estimate of phase, based on some criteria, and then use phase rather than time as the independent variable [18,19]. This approach can be useful, but limits one's ability to apply JIO and short-latency inference.

### 5.3. Use of continuous perturbations

Our approach to studying rhythmic movement depends on the ability to efficiently estimate $\phi$IRFs from experimental responses to perturbations. To estimate $\phi$IRFs efficiently, one must deal with two properties of neuromechanical systems: intrinsic variability and potentially different responses to natural vs. unnatural perturbations.

With a biological oscillator with intrinsic variability, using small perturbations to identify the $\phi$IRF may be difficult because the responses may be buried in the "noise". Increasing the amplitude of the perturbations to create a larger response may not be feasible because doing so might substantially change the nature of the response due to nonlinearities of the system. The alternative is to apply the perturbation many times for each perturbation phase, allowing sufficient time between perturbations for transients to decay, and then averaging the responses to estimate the $\phi$IRF. This approach may require prohibitively long experimental time to obtain accurate results. For perturbations of a fixed-point system rather than a limit-cycle system, there is a well-known solution to increasing the experimental efficiency of identifying the input-output mapping. Rather than applying spaced discrete perturbations, one applies broad-band perturbation signals. If the perturbation signal is close to white-noise, then the LTI IRF can be directly estimated in the time domain using the cross-covariance function between the input and output signals. However, white-noise perturbations of a biological oscillator may produce responses that are qualitatively different from those produced by more natural auto-correlated perturbation signals. When using auto-correlated perturbation signals, one common approach is to first characterize the input-output mapping in the frequency domain using a frequency response function (FRF) and then to convert to the time domain to obtain the IRF. Our use of HTFs to characterize responses in the frequency domain and then converting to $\phi$IRFs in the time domain in an extension of this approach.

## 5.4. Summary

In this paper we described a general approach to understanding the relationships among the open- and closed-loop properties of neuromechanical systems that control rhythmic behaviors such as walking. Our approach is an extension of methods widely used to study neuromechanical control systems that are approximately time invariant (LTI) [17] such as standing. Our framework is an extension of the theory of linear time period (LTP) systems [11–15] to limit cycle (LC) systems. Here we have applied this novel approach to the motor behavior of human treadmill walking, demonstrated its usefulness in learning about the control of that behavior and have described the powerful benefits provided by using this approach to investigate LC systems.

## Appendix

### A.1. The $\phi$IRF as a combination phase and transient $\phi$IRFs

We consider a vector $u(t)$ of small perturbations of an exponentially stable limit-cycle oscillator, where $t$ is time. The first component $u_1(t)$ of $u(t)$ is the experimental perturbation whose effect we wish to characterize. The other components of $u(t)$ are other experimental perturbations and intrinsic stochastic perturbations. Let $y(t)$ be a scalar output variable. Our goal is to non-parametrically characterize the mapping from $u_1(t)$ to $y(t)$.

By scaling time, we will assume that the unperturbed oscillator has frequency 1. Let $\theta(t)$ be an approximation of absolute phase based on $y(t)$ and other observed outputs of the system. We assume that $\theta(t)$ is continuously differentiable, monotonically increasing in time $t$, coincides with true absolute phase on the unperturbed limit cycle, and is causal, that is, $\theta(t)$ only depends on observations up to time $t$. Then the mapping from the perturbation vector $u(t)$ to $y(t)$ is described by a model of the form:

$$\dot{\theta}(t) = 1 + A_\theta(\theta(t))r(t) + g_\theta(\theta(t))u(t) + O(\|(r(t), u(t))\|^2), \qquad (9a)$$
$$\dot{r}(t) = \quad A_r(\theta(t))r(t) + g_r(\theta(t))u(t) + O(\|(r(t), u(t))\|^2), \qquad (9b)$$
$$y(t) = y_0(\theta(t)) + b(\theta(t))r(t) + O(\|r(t)\|^2), \qquad (9c)$$

where the vector $r$ describes deviations away from the limit cycle, $y_0(t)$ is the unperturbed waveform of $y(t)$, and all functions of $\theta$ are periodic with a period of 1.

We perform a perturbation analysis by letting

$$\begin{aligned}
\theta(t) &= \theta^{(0)}(t) + \varepsilon\theta^{(1)}(t) + O(\varepsilon^2), \\
r(t) &= \varepsilon r^{(1)}(t) + O(\varepsilon^2), \\
u(t) &= \varepsilon u^{(1)}(t) + O(\varepsilon^2), \\
y(t) &= y^{(0)}(t) + \varepsilon y^{(1)}(t) + O(\varepsilon^2),
\end{aligned}$$

where $\varepsilon$ is a small parameter that describes the size of perturbations. Then

$$\begin{aligned}
\dot{\theta}^{(0)}(t) + \varepsilon\dot{\theta}^{(1)}(t) &= 1 + \varepsilon A_\theta(\theta^{(0)}(t))r^{(1)}(t) + \varepsilon g_\theta(\theta^{(0)}(t))u^{(1)}(t) + O(\varepsilon^2), \\
\varepsilon\dot{r}^{(1)}(t) &= \varepsilon A_r(\theta^{(0)}(t))r^{(1)}(t) + \varepsilon g_r(\theta^{(0)}(t))u^{(1)}(t) + O(\varepsilon^2), \\
y^{(0)}(t) + \varepsilon y^{(1)}(t) &= y_0(\theta^{(0)}(t)) + \varepsilon y_0'(\theta^{(0)}(t))\theta^{(1)}(t) + \varepsilon b(\theta^{(0)}(t))r^{(1)}(t) + O(\varepsilon^2).
\end{aligned}$$

From the first of these equations, we have that $\dot{\theta}^{(0)}(t) = 1$. By shifting time, we will assume that $\theta^{(0)}(0) = 0$ so that $\theta^{(0)}(t) = t$. Then

$$\begin{align}
y^{(0)}(t) &= y_0(t), \tag{10a} \\
\dot{\theta}^{(1)}(t) &= A_\theta(t)r^{(1)}(t) + g_\theta(t)u^{(1)}(t), \tag{10b} \\
\dot{r}^{(1)}(t) &= A_r(t)r^{(1)}(t) + g_r(t)u^{(1)}(t), \tag{10c} \\
y^{(1)}(t) &= y_0'(t)\theta^{(1)}(t) + b(t)r^{(1)}(t). \tag{10d}
\end{align}$$

Note from (10b) and (10c) that the mappings from $u_1^{(1)}(t)$ to $\theta^{(1)}(t)$ and $r^{(1)}(t)$ are LTP with period 1 and can therefore can be characterized by $\phi$IRFs $h_\theta(t_r, t_p)$ and $h_r(t_r, t_p)$, respectively. From (10d) it follows that the mapping from $u_1^{(1)}(t)$ to $y^{(1)}(t)$ is also LTP with $\phi$IRF $h_y(t_r, t_p)$:

$$y^{(1)}(t_r) = \int_{-\infty}^{t_r} h_y(t_r, t_{p[})u_1^{(1)}(t_p)dt_s, \quad h_y(t_r, t_p) = y_0'(t_r)h_\theta(t_r, t_p) + b(t_r)h_r(t_r, t_p). \tag{11}$$

Note that the $\phi$IRF $h_y(t_r, t_p)$ is the sum of two components: $y_0'(t_r)h_\theta(t_r, t_p)$ describing phase resetting and $b(t_r)h_r(t_r, t_p)$ describing transients. Each of these components depend on the method of computing the phase estimate $\theta(t)$, but the their sum does not, since the $\phi$IRF $h_y(t_r, t_p)$ from $u_1^{(1)}(t)$ to $y^{(1)}(t)$ is defined independent of any particular phase estimate.

The LTP mapping (11) provides the first-order term of the mapping from the small perturbation $u_1(t)$ to the output $y(t)$. However, computing the $\phi$IRF $h_y(t_r, t_p)$ directly from data is problematic because system (9) is only approximately periodic in time $t$. To solve this problem, we define a transformed system in which approximate phase takes the place of time as the independent variable. We use the symbol $\vartheta$ to denote approximate phase as an independent variable to distinguish it from $\theta$, which is the function that maps time $t$ to approximate phase. Since $\theta(t)$ is a monotonically increasing function of time $t$, we can define its inverse $p: p(\theta(t)) = t$ and $\theta(p(\vartheta)) = \vartheta$. Now we can define $\tilde{r}(\vartheta) = r(p(\vartheta))$, $\tilde{u}(\vartheta) = u(p(\vartheta))$, $\tilde{y}(\vartheta) = y(p(\vartheta))$, and $\tilde{q}(\vartheta) = \dot{\theta}(p(\vartheta))$. With these definitions, we define the transformed system as

$$\tilde{q}(\vartheta) = 1 + A_\theta(\vartheta)\tilde{r}(\vartheta) + g_\theta(\vartheta)\tilde{u}(\vartheta) + O(\|(\tilde{r}(\vartheta), \tilde{u}(\vartheta))\|^2), \quad (12a)$$

$$\tilde{r}'(\vartheta) = \frac{\dot{r}(p(\vartheta))}{\dot{\theta}(p(\vartheta))} = A_r(\vartheta)\tilde{r}(\vartheta) + g_r(\vartheta)\tilde{u}(\vartheta) + O(\|(\tilde{r}(\vartheta), \tilde{u}(\vartheta))\|^2), \quad (12b)$$

$$\tilde{y}(\vartheta) = y_0(\vartheta) + b(\vartheta)\tilde{r}(\vartheta) + O(\|\tilde{r}(\vartheta)\|^2), \quad (12c)$$

Note that the transformed system (12) is similar to the original system (9), but with two important differences. First, the right-hand sides of (9) are not periodic in time $t$, whereas the right-hand sides of (12) are periodic in approximate phase $\vartheta$. Second, (9a) describes phase dynamics that affect the transient dynamics (9b), whereas (12a) describes an output variable $\tilde{q}(\vartheta)$ that does not affect the transient dynamics (12b).

Now we can carry out a perturbation analysis of system (12) similar to our previous perturbation analysis of system (9) by letting

$$\begin{aligned}
\tilde{q}(\vartheta) &= 1 + \varepsilon \tilde{q}^{(1)}(\vartheta) + O(\varepsilon^2), \\
\tilde{r}(\vartheta) &= \varepsilon \tilde{r}^{(1)}(\vartheta) + O(\varepsilon^2), \\
\tilde{u}(\vartheta) &= \varepsilon \tilde{u}^{(1)}(\vartheta) + O(\varepsilon^2), \\
\tilde{y}(\vartheta) &= y_0(\vartheta) + \varepsilon \tilde{y}^{(1)}(\vartheta) + O(\varepsilon^2),
\end{aligned}$$

leading to the first-order effects

$$\tilde{q}^{(1)}(\vartheta) = A_\theta(\vartheta)\tilde{r}^{(1)}(\vartheta) + g_\theta(\vartheta)\tilde{u}^{(1)}(\vartheta), \quad (13a)$$

$$\tilde{r}^{(1)'}(\vartheta) = A_r(\vartheta)\tilde{r}^{(1)}(\vartheta) + g_r(\vartheta)\tilde{u}^{(1)}(\vartheta), \quad (13b)$$

$$\tilde{y}^{(1)}(\vartheta) = b(\vartheta)\tilde{r}^{(1)}(\vartheta). \quad (13c)$$

Comparing (10b) and (13b), we see that the LTP map from $\tilde{u}_1^{(1)}(\vartheta)$ to $\tilde{r}^{(1)}(\vartheta)$ is the same as the map from $u_1^{(1)}(t)$ to $r^{(1)}(t)$ and thus has the $\phi$IRF $h_r(\vartheta_r, \vartheta_p)$. Now, from (13c) we have that the $\phi$IRF $\tilde{h}_y(\vartheta_r, \vartheta_p)$ from $\tilde{u}_1^{(1)}(\vartheta)$ to $\tilde{y}^{(1)}(\vartheta)$ is

$$\tilde{h}_y(\vartheta_r, \vartheta_p) = b(\vartheta_r) h_r(\vartheta_r, \vartheta_p). \quad (14)$$

Note that (14) is analogous to (11), except that the phase-resetting component of $\tilde{h}(\vartheta_r, \vartheta_s)$ is missing.

To compute the phase resetting component, we compare (10a) and (13a) and note that the LTP map from $\tilde{u}_1^{(1)}(\vartheta)$ to $\tilde{q}^{(1)}(\vartheta)$ is the same as the map from $u_1^{(1)}(t)$ to $\dot{\theta}^{(1)}(t)$. Therefore, if $\tilde{h}_q(\vartheta_r, \vartheta_p)$ is the $\phi$IRF from $\tilde{u}_1^{(1)}(\vartheta)$ to $\tilde{q}^{(1)}(\vartheta)$, then the $\phi$IRF $h_\theta(t_r, t_p)$ from $u_1^{(1)}(t)$ to $\theta^{(1)}(t)$ is given by

$$h_\theta(t_r, t_p) = \int_{t_s}^{t_r} \tilde{h}_q(t, t_p) dt. \quad (15)$$

Putting these pieces together, to compute the $\phi$IRF $h_y(t_r, t_p)$ from $u_1^{(1)}(t)$ to $y^{(1)}(t)$ we perform the following steps:

1. Let approximate phase $\vartheta$ take the place of time $t = p(\vartheta)$ as the independent variable and compute $\tilde{u}_1(\vartheta) = u_1(p(\vartheta))$, $\tilde{y}(\vartheta) = y(p(\vartheta))$, and $\tilde{q}(\vartheta) = \dot{\theta}(p(\vartheta))$.

2. Approximate the unperturbed waveform $y_0(\vartheta)$ as the mean of $\tilde{y}(\vartheta)$ as a function of $\vartheta$.

3. Compute the approximations $\tilde{u}_1^{(1)}(\vartheta) = \tilde{u}_1(\vartheta)/\varepsilon$, $\tilde{y}^{(1)}(\vartheta) = (\tilde{y}(\vartheta) - y_0(\vartheta))/\varepsilon$ and $\tilde{q}^{(1)}(\vartheta) = (\tilde{q}(\vartheta) - 1)/\varepsilon$. (The value used for $\varepsilon$ is arbitrary, since the result of the next step only depends on the relative sizes of $\tilde{u}_1^{(1)}(\vartheta)$, $\tilde{y}^{(1)}(\vartheta)$ and $\tilde{q}^{(1)}(\vartheta)$.)

4. Compute the $\phi$IRF $\tilde{h}_y(\vartheta_r, \vartheta_s)$ from $\tilde{u}_1^{(1)}(\vartheta)$ to $\tilde{y}^{(1)}(\vartheta)$ and the $\phi$IRF $\tilde{h}_q(\vartheta_r, \vartheta_p)$ from $\tilde{u}_1^{(1)}(\vartheta)$ to $\tilde{q}^{(1)}(\vartheta)$.

5. Use (15) to compute the $\phi$IRF $h_\theta(t_r, t_p)$ from $u_1^{(1)}(t)$ to $\theta^{(1)}(t)$.

6. Compute

$$h_y(t_r, t_p) = \tilde{h}_y(t_r, t_p) + y_0'(t_r) h_\theta(t_r, t_p). \tag{16}$$

*Note*: Recall that we assumed that time $t$ was scaled so that the unperturbed oscillator has frequency 1 and shifted so that $t = 0$ corresponds to $\vartheta = 0$. Scaling time consists of multiplying time by $f_0$, the original oscillator frequency in Hz. Scaling time has no effect on $\tilde{u}_1(\vartheta)$ and $\tilde{y}(\vartheta)$ and scales $\tilde{q}(\vartheta) = \dot{\theta}(p(\vartheta))$ by $1/f_0$. Therefore, we can perform the above analysis without scaling time simply by redefining $\tilde{q}(\vartheta)$ in step 2 to be $\tilde{q}(\vartheta) = \dot{\theta}(p(\vartheta))/f_0$, where $f_0$ is approximated by the average value of $\dot{\theta}(p(\vartheta))$. In this case, we should still interpret $t_r$ and $t_s$ in (16) as normalized times relative to a phase-0 event.

Due to the approximation in step 3 above, our estimate of $h_y(t_r, t_p)$ is correct to 0-th order in $\varepsilon$ with errors of $O(\varepsilon)$ that depend on factors such as the method used to compute approximate phase $\theta(t)$. There will be additional errors due to estimating $f_0$, $y_0(\vartheta)$, $\tilde{h}_y(\vartheta_r, \vartheta_p)$, and $\tilde{h}_q(\vartheta_r, \vartheta_p)$ from a finite amount of data.

### A.2. Using HTFs to compute $\phi$IRFs

We now consider how to use data to estimate the $\phi$IRFs from step 4 above. To simplify notation, we drop the tildes, subscripts and superscripts and consider how to estimate the $\phi$IRF from an scalar input $u(\vartheta)$ to a scalar output $y(\vartheta)$ that are functions of estimated phase $\vartheta$ (mod 1) for $0 \le \vartheta \le N_c$, where $N_c$ is the number of cycles of data. We define the Fourier transforms of a single window of data as

$$U_k(f) = \int_k^{k+n_c} w(\vartheta - k) u(\vartheta) e^{-2\pi i \vartheta} d\vartheta, \quad Y_k(f) = \int_k^{k+n_c} w(\vartheta - k) y(\vartheta) e^{-2\pi i \vartheta} d\vartheta,$$

where $n_c$ is an even integer specifying the number of cycles in a window, the integer $k$ is the starting phase of the window, $w$ is a tapered window function to reduce side-lobe leakage [1], and $f$ is normalized frequency such that $f = 1$ corresponds to the oscillation frequency $f_0$. Here we choose the Hanning window for $w$: $w(x) = (1 - \cos(2\pi x/n_c))/2$ and use windows with 50% overlap. Then the number of overlapping window is $n_w$, the greatest integer less than or equal to $2N_c/n_c - 1$. We estimate the power spectral density (PSD) $p_{uu}(f_1)$ of the input and the double-frequency cross-spectral density (CSD) $p_{uy}(f_1, f_2)$ between the input and output as

$$p_{uu}(f_1) = \frac{1}{n_w W} \sum_{k=0}^{n_w-1} |U_{kn_c/2}(f_1)|^2, \quad p_{uy}(f_1, f_2) = \frac{1}{n_w W} \sum_{k=0}^{n_w-1} U^*_{kn_c/2}(f_1) Y_{kn_c/2}(f_2),$$

where $f_1$ is normalized input frequency, $f_2$ is normalized output frequency, $W$ is the mean-squared value of $w(x)$, and the asterisk '*' denotes complex conjugation. Using the PSD and CSD, we compute the double-frequency frequency response function (FRF) as $H(f_1, f_2) = p_{uy}(f_1, f_2)/p_{uu}(f_1)$. The $k$-th mode of the HTF is defined as $H_k(f_1) = H(f_1, f_1 + k)$. $H_k(f_2)$ describes how input at normalized frequency $f_1$ affects output at normalized frequency $f_2 = f_1 + k$. Finally, the $\phi$IRF is computed as the inverse two-dimensional Fourier transform of the HTF:

$$h_y(\vartheta_r, \vartheta_p) = \int_{-\infty}^{\infty} \sum_{k=-\infty}^{\infty} H_k(f_1) e^{2\pi i [k\vartheta_p + f_1(\vartheta_r - \vartheta_p)]} df_1$$

**Acknowledgements.** Work supported by NSF grant BCS-1230311. We thank Noah Cowan and Eric Tytell for helpful discussions related to this work.

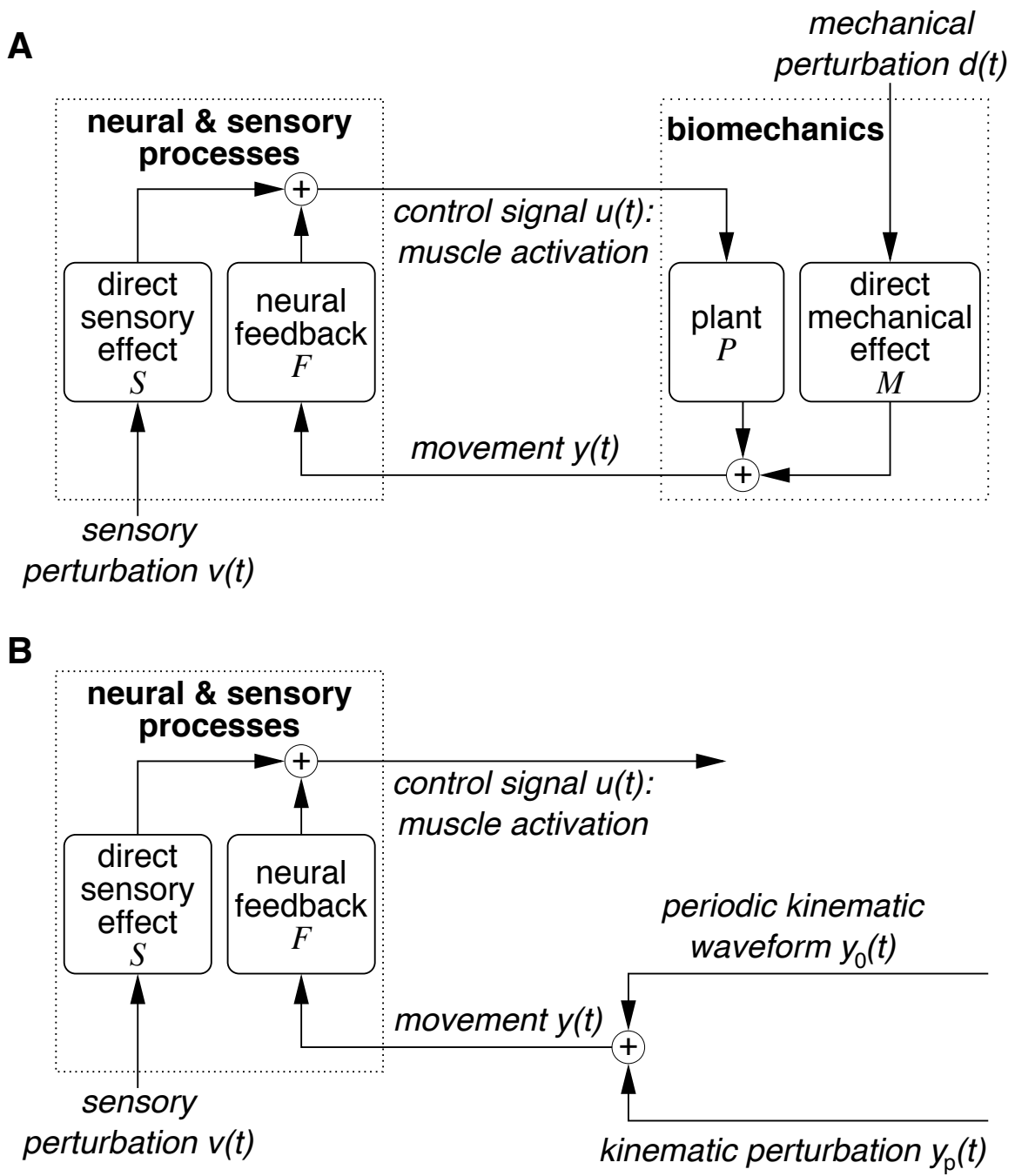

*Figure 1:* Schematic diagram of the neural control of rhythmic movement with weak sensory and mechanical perturbations. A: The closed-loop system in which muscle activation causes movement and movement causes muscle activation. Muscle activation is the control signal; experimentally, it is measured using electromyography (EMG). The variables $u(t)$, $y(t)$, $v(t)$ and $d(t)$ can be either a scalars or a vectors. B: A hypothetical open-loop condition in which movement affects the control signal, but the control signal does not affect movement. Diagrams A and B also describe the neural control of posture, except that the periodic kinematic waveform $y_0(t)$ in B becomes a constant.

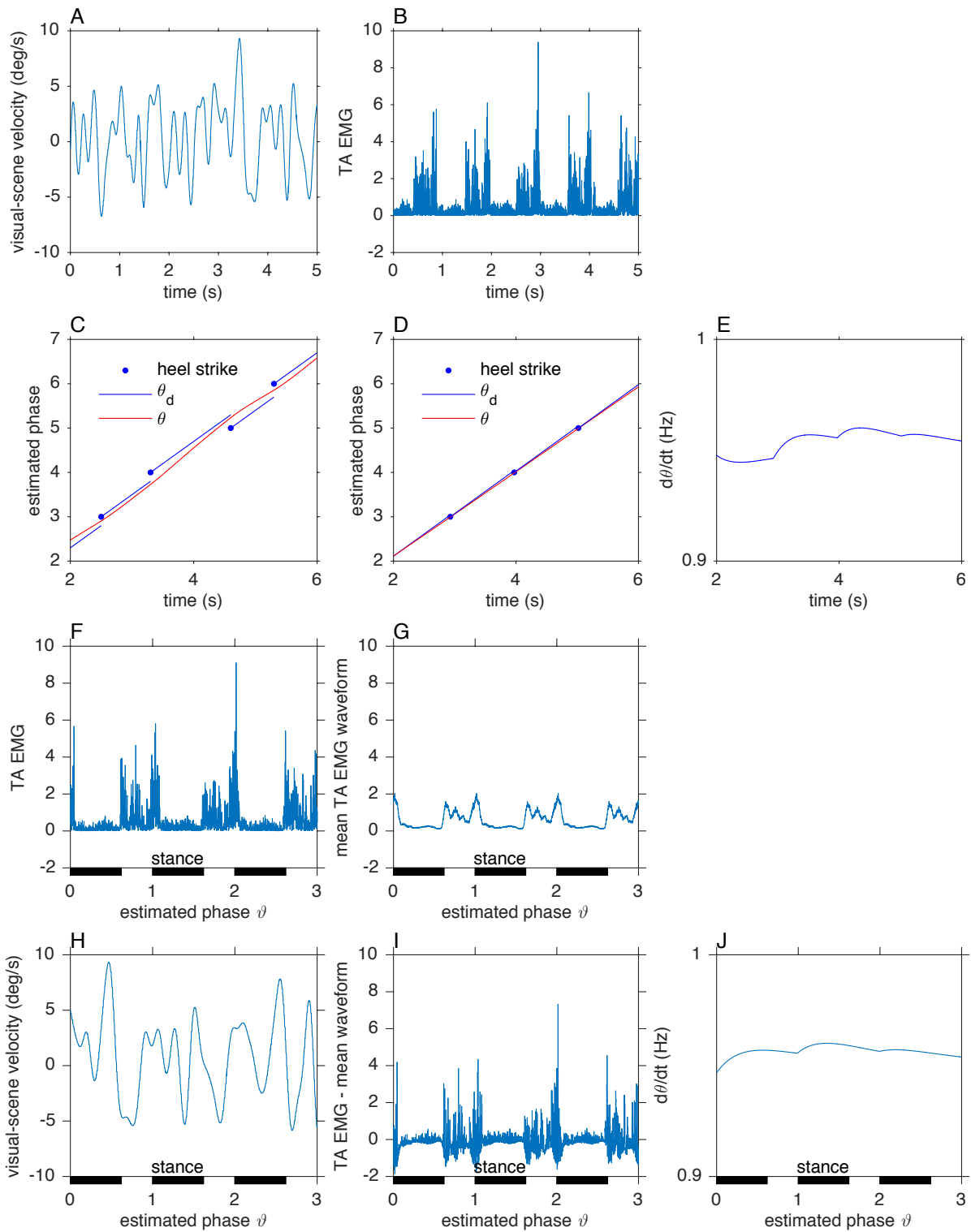

*Figure 2:* Processing of signals prior to spectral analysis. A: Original perturbation signal, the visual scene velocity. B: Original output signal, the rectified EMG signal recorded from the tibialis anterior muscle. C: Estimation of phase illustrated with exaggerated variation in cycle periods. D: Estimation of phase based on actual heel-strike times. E: Time derivative of estimated phase. F: Output signal as a function of estimated phase. G: Mean periodic waveform of output signal. H: Perturbation signal as a function of estimated phase. I: Deviation of output signal from its mean periodic waveform as a function of estimated phase. J: Time derivative estimated phase as a function of estimated phase.

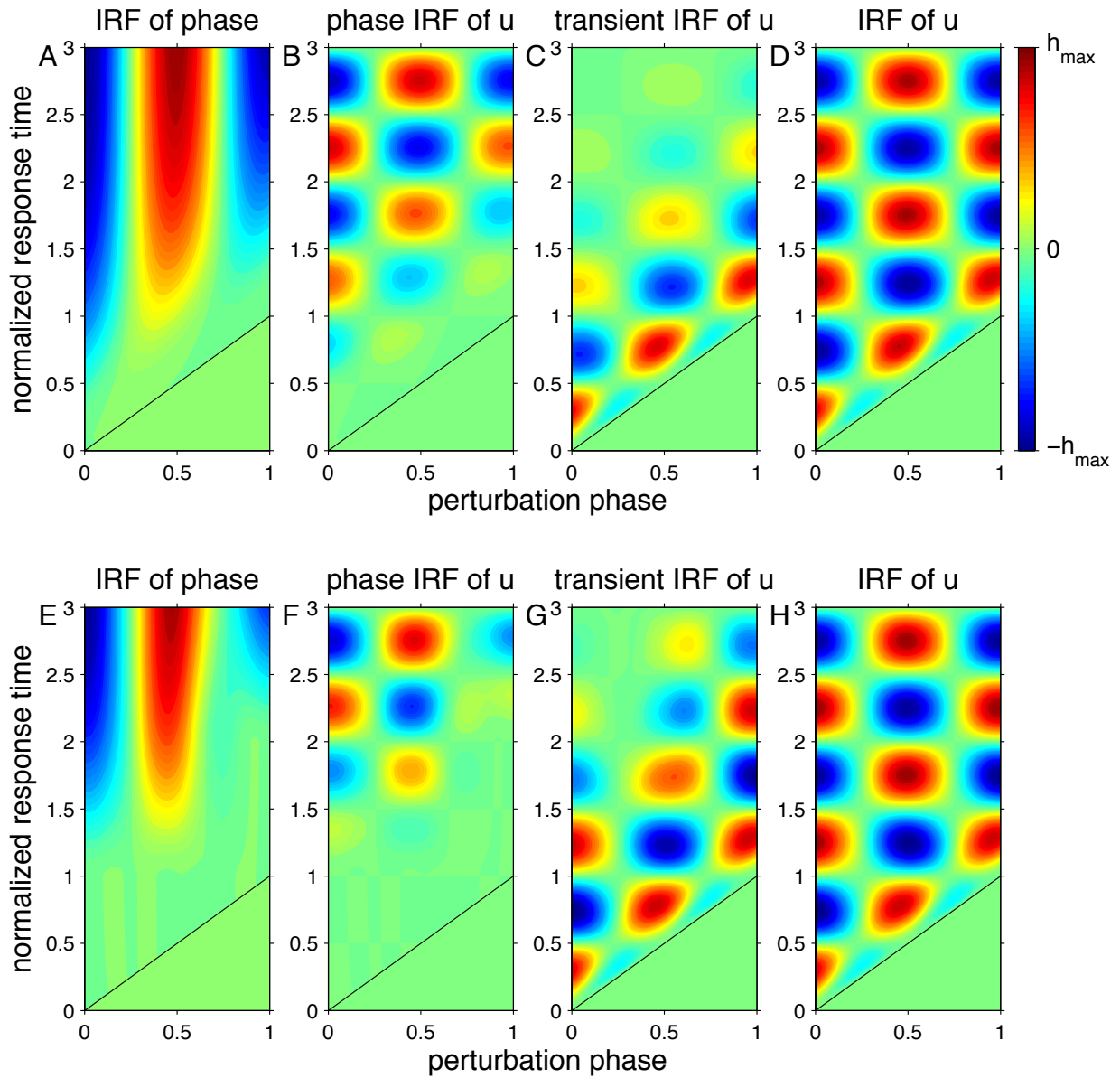

*Figure 3:* Method of computing closed-loop IRFs applied to simulated data. The input is the sensory perturbation $v(t)$ and the output is the control signal $u(t)$. A-D shows IRFs for estimated phase obtained by filtering phase $\phi_f(t)$ in the model. E-G shows IRFs for estimated phase based on event times. The similarity of the IRFs in D and H illustrates that, to first order in perturbation size, the computed IRF does not depend on the method used to estimate phase. Values of $h_{max}$ are 0.40 s$^{-1}$ in A and E and 2.73 s$^{-1}$ otherwise.

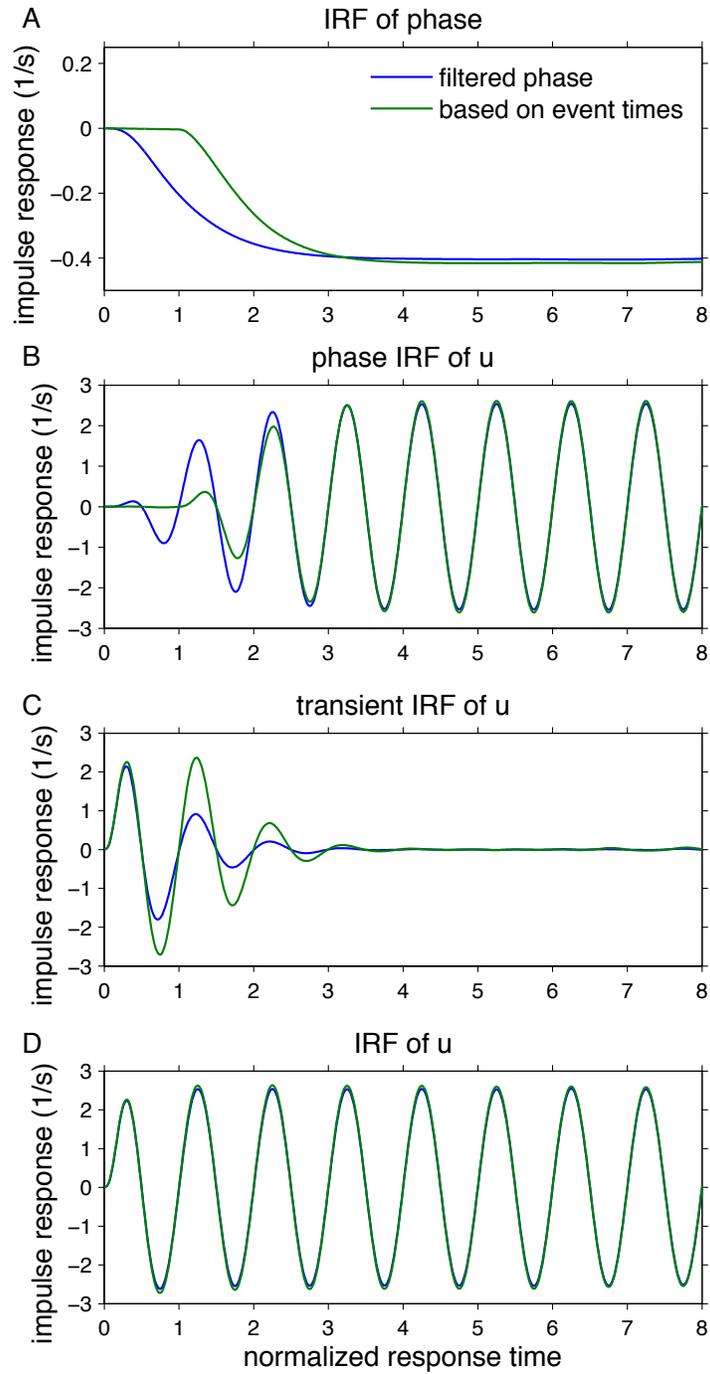

*Figure 4:* Vertical slices through the IRFs of Fig. 1 at stimulus phase 0. The range of normalized response time has been increased to illustrate that both IRFs of phase in A converge to the same value.

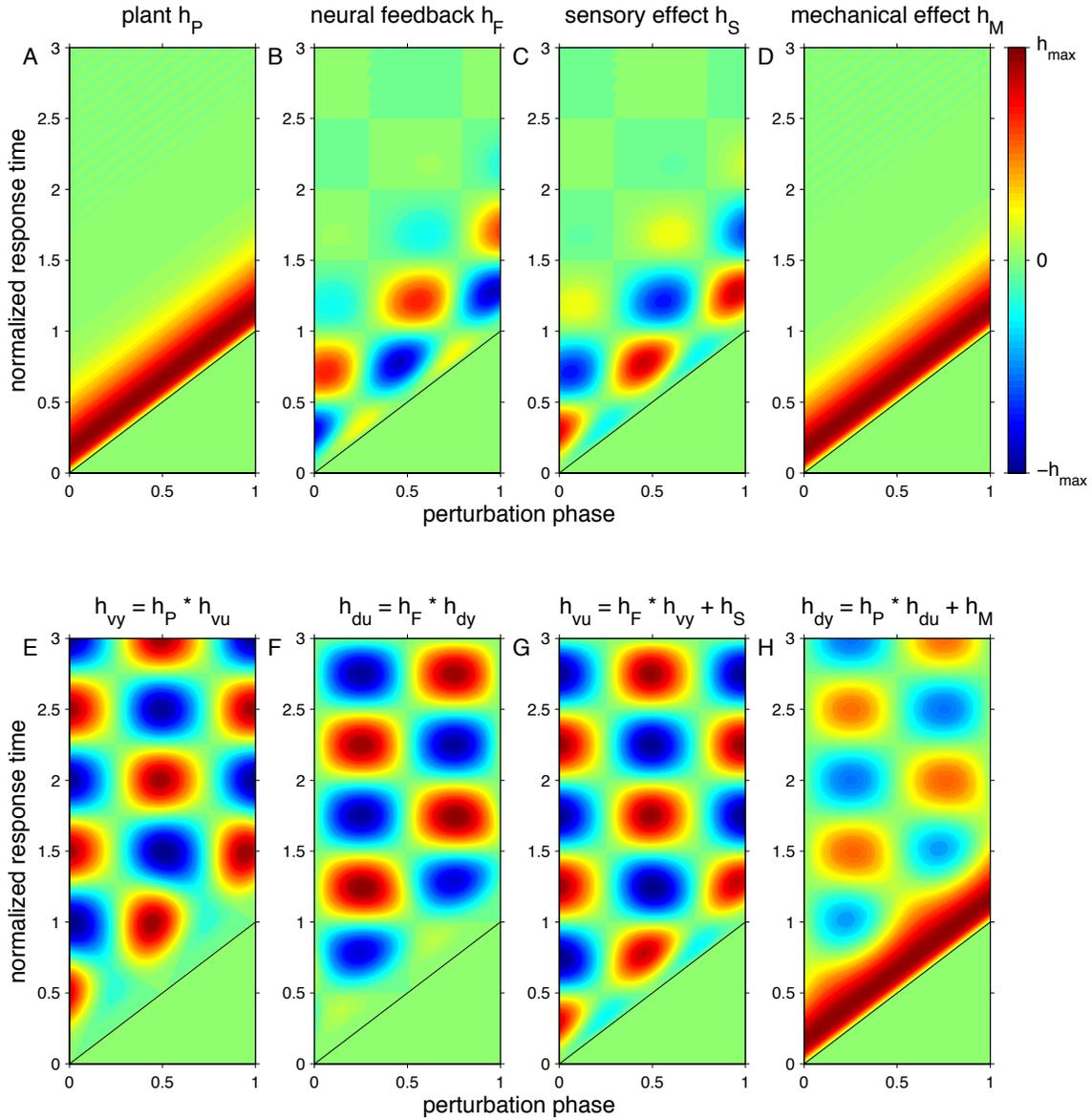

*Figure 5:* A–D: Open-loop model IRFs. E–H: Closed-loop model IRFs predicted by open-loop IRFs. Values of $h_{max}$ are: 4.69 s$^{-1}$ in A, D and H; 2.72 s$^{-1}$ in B, C and G; 2.77 s$^{-1}$ in E; and 2.65 s$^{-1}$ in F.

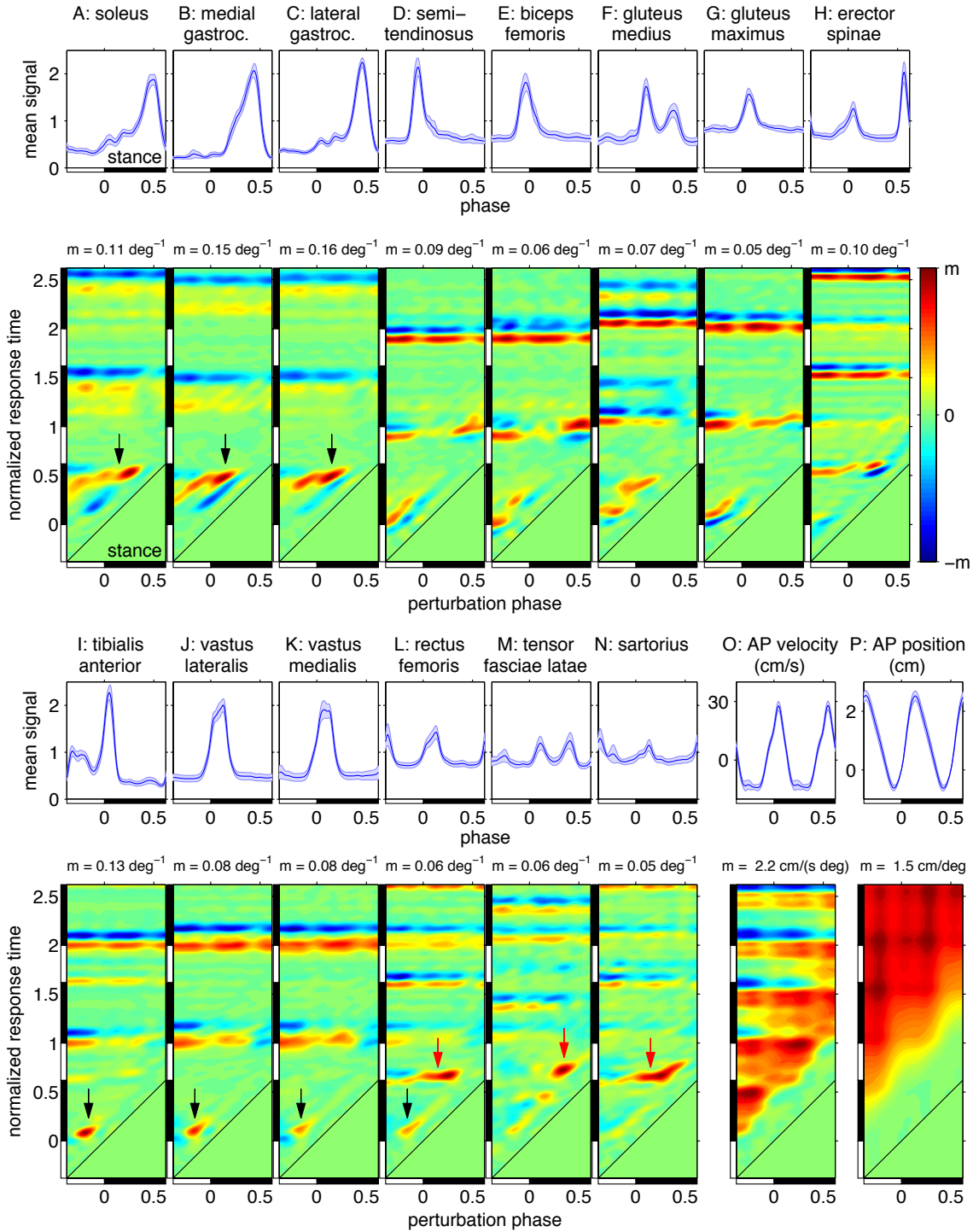

*Figure 6:* Closed-loop experimental IRFs showing responses to visual-scene velocity. Responses are shown for normalized EMG from 14 muscles (A–N). These muscle-activation responses led to changes in walking speed (O) and position on the treadmill (P), computed using the anterior-posterior coordinate of the midpoint of the two hip markers. Black bars indicate values of 0 to 0.62 of stimulus phase or normalized response time corresponding to stance for the reference leg. Above each IRF is the mean of the response signal with errors bars indicating 95% confidence intervals based on *t*-tests.